# Thanking the World: Exploring Gender-Based Differences in Acknowledgment Patterns and Support Systems in Theses

Manika Lamba, University of Illinois Urbana-Champaign
Hendrik Erz, Linköping University

**Abstract:** Research on acknowledgment sections of scientific papers has gained significant attention, but there remains a dearth of studies examining acknowledgments in the context of Electronic Theses and Dissertations (ETDs). This paper addresses this gap by investigating the sources of support for male and female researchers in completing their master's or doctoral theses, focusing on the discipline of Library and Information Science (LIS). We utilize a novel method of extracting the various types of support systems that are acknowledged in 1,252 ETDs using RoBERTa-based models. The most prominent forms of support acknowledged by researchers are academic, moral, financial, and religious support. While there are no significant gender-based differences in religious and financial support, the ratio of academic to moral support acknowledged by researchers shows strong gender-based variation. Additionally, advisors display a preference for supervising same-gender researchers. By comprehending the nuances of support systems and the unique challenges faced by researchers of different genders, we can foster a more inclusive and supportive academic environment. The insights gained from this research have implications for improving mentoring practices and promoting gender equality in academia.

## 1.1 Introduction

Research on acknowledgment sections of scientific papers has grown substantially since the nineties. In one of the first analyses on the subject, Cronin et al. [1] found a positive correlation between the number of acknowledgments in a paper and the citations it receives in the future. There are many papers studying acknowledgment as a reward system [2], but there is no consensus as to its value and functions [3]. Acknowledgments have been classified as contributions, disclaimers, and authorial voices [4]. The aforementioned studies particularly use research articles as their main source of data. One of the main issues in acknowledgment research is data availability [5]. This may partially explain the lack of publications analyzing acknowledgment sections from ETDs [6]. Specifically, it has been difficult to work with large textual corpora in the past [7]. Against the backdrop of the growth of text analysis methods in the past few years in the social sciences, we show how the use of Natural Language Processing (NLP) with BERT models to extract and analyze acknowledgment sections can help researchers make sense of large text corpora.

A notable gap exists in the examination of acknowledgments within the context of Electronic Theses and Dissertations (ETDs), particularly in the field of Library and Information Science (LIS). Our study is the first one to analyze acknowledgment sections in LIS ETDs spanning 94 years using computational text analysis methods.

This paper aims to address this gap by investigating the sources of support for male and female researchers in completing their master's or doctoral theses[1]. Our research goes beyond traditional acknowledgment analysis by categorizing the types of support systems mentioned. The broad coding scheme ensures a comprehensive exploration of the acknowledgment landscape while allowing us to pinpoint gender-based differences in acknowledgment patterns to understand the unique challenges and support systems faced by male and female researchers. As indicated by Yang et al. [8], the graduate school network plays a crucial role in predicting the attainment of leadership positions for both male and female students where high-placing women exhibit distinctive networks. In a related context, Shin [9] demonstrated that women working under female supervisors tend to benefit from increased family and organizational support. Moreover, Mason et al. [10] shed light on the challenges faced by doctoral researchers who are also mothers. The study highlights how the responsibilities of motherhood can significantly strain these researchers, impacting their capacity to conduct and document their research. These challenges are exacerbated by unequal and gendered role expectations, limited resources, and insufficient support, often at the expense of the physical and mental well-being of doctoral researchers. Adding to this perspective, Cidlinská and Zilincikova [11] delve into the disparities in academic careers between men and women. They discuss how women's academic trajectories tend to be shorter, progress more slowly, and encounter more hurdles compared to their male counterparts. The study also explores gender differences in contemplating leaving an academic career, examining this phenomenon across various stages of individuals' professional journeys.

The major research questions of this study are:

RQ1 What are the important support systems for male and female researchers to finish their master's or doctoral research?

RQ2 What are the major sentiments that researchers felt throughout their journey of research?

The significance of this study lies in shedding light on the performance of academic support systems as indicated by the acknowledged types of support. By uncovering the nuances of support systems and gender dynamics, we aim to contribute insights that can inform mentoring practices and promote gender equality in academia. Thus, this research not only addresses the existing gap in acknowledgment studies but also pioneers the application of NLP to analyze ETDs. The focus on gender-related aspects adds depth to our understanding of acknowledgment patterns, providing valuable contributions to the broader discourse on support systems and gender dynamics in academic pursuits.

## 1.2 Material and methods

We analyzed 1,252 ETDs in the English language from the ProQuest Dissertations and Theses (PQDT) database. To extract these, we used the query *SU.exact("LIBRARY SCIENCE")* and selected theses from 1927 to 2020. This yielded over 4,000 ETDs, which we downloaded manually between late 2019 and 2021[2]. We performed optical character recognition (OCR) with the program Tesseract on the PDF files to extract the text layer for all files. As we were only interested in the acknowledgment sections, we had to extract those sections, which proved to be difficult. We first attempted to extract the acknowledgment sections using a script that utilized simple heuristics. In the end, the script could extract acknowledgment sections from about one-third of the theses. The variance in styles regarding the acknowledgment sections made it hard for the script to extract all sections properly. Therefore, we extracted further sections manually. After extraction, the data collection phase ended with 1,252 acknowledgment sections or 20,834 individual sentences. We further reduced this corpus to only include theses from 1960 to 2020, since between 1927 and 1959 the number of theses was too small, and we could not determine the gender of researchers and advisors in this period.

---

[1] We use "theses" to refer to both master and doctoral research projects to avoid ambiguity since some countries/universities use "theses" to refer to masters' projects whereas others use them for doctoral projects.

The next step was to classify the acknowledgment sections according to the support systems that are being acknowledged. We constructed a coding guide of eight types of "support" that we expected to find in the theses (Table 1). This list is inspired by what Paul-Hus and Desrochers [4] used to classify research articles.

Besides classifying the acknowledgment sections according to types of support, we were also interested in the sentiment of these sections. We labeled each sentence as *"very positive"*, *"somewhat positive"*, or *"neutral"*[3].

**Table 1: Codebook: Categories of Acknowledgment Content**

| *Category* | *Definition* |
|---|---|
| **Academic Support** | It involves guidance from Professors or Doctors who, as supervisors or committee members, provide crucial assistance to students or researchers. Acknowledgment is typically done using their full names and academic titles. |
| **Moral Support** | It encompasses encouragement and comfort provided by colleagues, friends, and family in private settings. This form of support is typically expressed using first names and serves to uplift and strengthen individuals emotionally during challenging times. |
| **Financial Support** | It refers to assistance provided by funding agencies or universities to enable individuals to complete their academic theses or projects. This support involves monetary contributions that cover various expenses associated with research, education, or specific project needs. |

---

[2] The PDF files were downloaded manually, as the database did not allow web scraping on their website and did not have an API.
[3] During the initial screening, we never found an acknowledgment sentence with a negative sentiment. After consultation with four coders, we have decided to focus on a positive sentiment scale ranging from "neutral" to "very positive".

| | | |
|---|---|---|
| | **Technical Support** | It involves assistance with specific skills or tasks related to writing, proofreading, and adopting new techniques. It includes guidance and help provided to individuals to enhance their technical abilities, refine written work, and acquire proficiency in novel methods or approaches. |
| | **Access to Data** | It refers to the provision of assistance or support in obtaining the necessary information or datasets required for a particular purpose. This support may involve facilitating the acquisition, retrieval, or permission to use specific data sets, enabling individuals or organizations to gather relevant information for analysis, research, or other purposes. |
| | **Religious Support** | It involves drawing on quotes from religious texts and making references to the names of deities or higher powers for inspiration, guidance, or comfort. This form of support often includes expressions rooted in faith and spirituality to provide strength and solace in various situations. |
| | **Library Support** | It refers to the assistance provided by libraries and librarians to help individuals complete searches, tasks, or research endeavors. This support may involve guidance on navigating library resources, locating relevant materials, and utilizing research tools effectively to enhance the quality and efficiency of the information retrieval process. |
| | **Other** | It refers to miscellaneous or additional information not covered by specific categories. In this context, it encompasses diverse sentences or details that do not fit into predefined categories. |

We performed the classification in a supervised setting, utilizing a large language model (LLM) for the task. First, we extracted a random sample of 903 sentences from the corpus and manually labeled them according to the type of support and sentiment[4]. This small "gold-standard" dataset comprised about 4% of our entire corpus. The aim was to apply the same labeling strategy to the rest of the corpus. For this, we utilized a simple sequence classification task. Research has shown that LLMs, such as BERT, are very good at learning the patterns that predict labels in a small dataset, and applying those labels to a large corpus, effectively copying the coding strategy of human annotators [12, 13].

---

[4] Four coders coded the sentences individually. Final categories for both support and sentiment labels were chosen if at least 3 coders chose the same label.

To label the entire corpus according to the labeling strategy, we trained two RoBERTa-base transformer models on our data [14] and had them label the remaining sentences within the corpus accordingly.

We trained two RoBERTa-base models; one for the support category classification task (Table 2), and one for the sentiment classification task (Table 3). To construct our training data, we randomly chose 80% of the sentences in our dataset to train our classifiers and kept 20% of the sentences to calculate our metrics ("held-out", or validation dataset). Both models were trained for 15 epochs with an adaptive learning rate (LR) of $5 \times 10^{-5}$ and an epsilon for the ADAM optimizer of $\epsilon = 10^{-8}$. After each epoch, we calculated the F1 scores for each individual category, which resulted in a set of eight F1 scores for the support classifier and three F1 scores for the sentiment classifier. After training, we selected the best model based on the average F1-score [12] and used this to classify our entire corpus of 20,834 sentences. Tables 2 and 3 below report all F1 and accuracy scores, with the best iteration emphasized.

**Table 2: Metrics for the Support Category Classifier**

| # | F1 (avg) | F1 (acad) | F1 (moral) | F1 (tech) | F1 (data) | F1 (lib) | F1 (fin) | F1 (rel) | F1 (other) | Accuracy |
|---|---|---|---|---|---|---|---|---|---|---|
| 1 | 0.22 | 0.76 | 0.71 | 0.0 | 0.0 | 0.0 | 0.0 | 0.0 | 0.27 | 0.65 |
| 2 | 0.27 | 0.78 | 0.77 | 0.0 | 0.0 | 0.0 | 0.0 | 0.0 | 0.58 | 0.70 |
| 3 | 0.36 | 0.79 | 0.78 | 0.0 | 0.0 | 0.0 | 0.0 | 0.67 | 0.64 | 0.72 |
| 4 | 0.38 | 0.78 | 0.73 | 0.0 | 0.0 | 0.25 | 0.0 | 0.67 | 0.59 | 0.69 |
| 5 | 0.5 | 0.82 | 0.68 | 0.0 | 0.0 | 0.6 | 0.55 | 0.67 | 0.68 | 0.72 |
| 6 | 0.5 | 0.85 | 0.71 | 0.25 | 0.0 | 0.33 | 0.55 | 0.67 | 0.62 | 0.70 |
| 7 | 0.57 | 0.82 | 0.70 | 0.33 | 0.33 | 0.33 | 0.67 | 0.67 | 0.68 | 0.71 |
| 8 | 0.57 | 0.85 | 0.72 | 0.4 | 0.4 | 0.31 | 0.6 | 0.67 | 0.6 | 0.72 |
| 9 | 0.59 | 0.83 | 0.70 | 0.29 | 0.4 | 0.44 | 0.75 | 0.67 | 0.62 | 0.72 |
| 10 | 0.58 | 0.83 | 0.71 | 0.29 | 0.4 | 0.55 | 0.6 | 0.67 | 0.60 | 0.71 |
| 11 | 0.55 | 0.80 | 0.67 | 0.25 | 0.33 | 0.36 | 0.67 | 0.67 | 0.64 | 0.69 |
| 12 | 0.51 | 0.81 | 0.68 | 0.0 | 0.33 | 0.33 | 0.67 | 0.67 | 0.60 | 0.69 |
| 13 | 0.53 | 0.81 | 0.69 | 0.0 | 0.33 | 0.46 | 0.67 | 0.67 | 0.63 | 0.69 |
| **14** | **0.60** | **0.81** | **0.68** | **0.44** | **0.4** | **0.46** | **0.67** | **0.67** | **0.68** | **0.71** |
| 15 | 0.60 | 0.80 | 0.68 | 0.44 | 0.4 | 0.46 | 0.67 | 0.67 | 0.64 | 0.70 |

*Note: The numbers are rounded. The categories used for the table are: academic, moral, technical, data access, library, financial, religious, and unknown/other support. The best model is number 14.*

Table 3: Metrics for the Sentiment Classifier

| # | F1 (average) | F1 (very positive) | F1 (somewhat positive) | F1 (neutral) | Accuracy |
|---|---|---|---|---|---|
| 1 | 0.36 | 0.48 | 0.45 | 0.15 | 0.44 |
| 2 | 0.43 | 0.36 | 0.78 | 0.16 | 0.66 |
| 3 | 0.43 | 0.04 | 0.77 | 0.47 | 0.64 |
| 4 | 0.58 | 0.51 | 0.76 | 0.45 | 0.67 |
| 5 | 0.48 | 0.21 | 0.76 | 0.46 | 0.64 |
| 6 | 0.55 | 0.45 | 0.77 | 0.44 | 0.67 |
| 7 | 0.58 | 0.49 | 0.76 | 0.49 | 0.67 |
| 8 | 0.57 | 0.51 | 0.74 | 0.46 | 0.65 |
| **9** | **0.59** | **0.52** | **0.75** | **0.49** | **0.66** |
| 10 | 0.57 | 0.52 | 0.76 | 0.41 | 0.67 |
| 11 | 0.56 | 0.5 | 0.71 | 0.47 | 0.63 |
| 12 | 0.57 | 0.51 | 0.74 | 0.46 | 0.65 |
| 13 | 0.56 | 0.45 | 0.75 | 0.49 | 0.66 |
| 14 | 0.55 | 0.45 | 0.74 | 0.47 | 0.65 |
| 15 | 0.56 | 0.46 | 0.74 | 0.47 | 0.65 |

*Note: The numbers are rounded. The best model is number 9.*

In order to analyze potential gender-based discrepancies, we additionally performed gender assignments to the names of supervisors and authors of the theses, utilizing a list of commonly given names[5] and gendered nouns collected from the data files[6] from Garg et al. [15].

**1.3 Results**

The results of our analysis reveal several noteworthy findings. First, we find that most sentences seem to be of *"neutral"* sentiment, followed by *"somewhat positive"* and lastly *"very positive"*. This lack of variation in sentiment suggests that the style of acknowledgments is very formalized across our corpus, requiring a certain style from all authors that is academic but allows for a more emotional tone in these sections[7].

The most frequently acknowledged support category was academic support, comprising 26.72% of the acknowledgments, followed by moral support at 11.22% and library support at 3.40% (Fig. 1). Notably, *"religious"* support did not play a significant role in the past, but increased in frequency in the 21st century. More than half of all sentences were categorized as *"other"* (53.93%), indicating that the support systems are more diverse than expected.

---

[5] https://www.geeksforgeeks.org/python-gender-identification-by-name-using-nltk/
[6] https://github.Com/nikhgarg/embeddingdynamicstereotypes/tree/master/data
[7] While this is a reasonable conclusion, it should be noted that the F1-scores for the "very positive" and "neutral" sentiment categories are comparatively small, and thus it could be that our classifier had difficulties discerning those categories. However, given that the F1 score of the "somewhat positive" category is high, we believe that our interpretation holds.

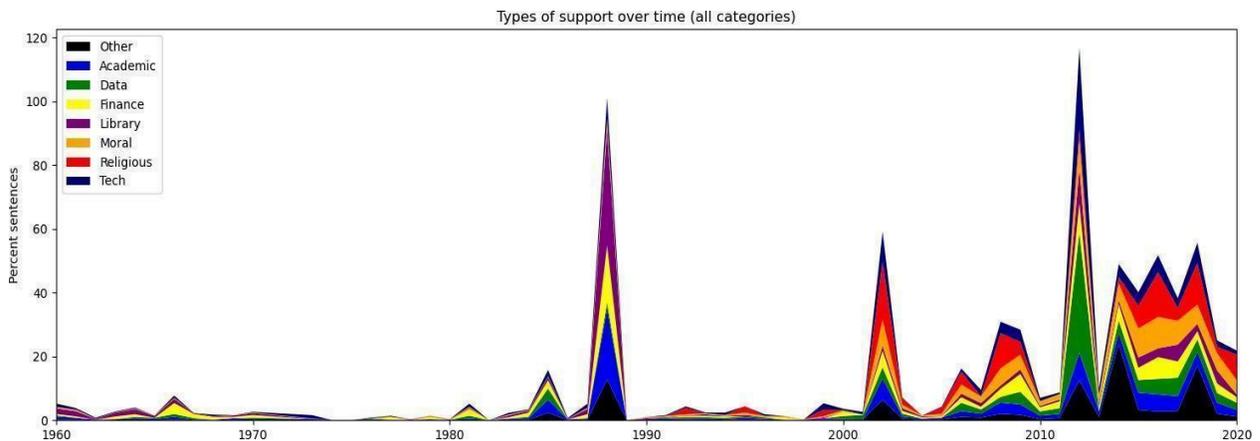

**Figure 1: Evolution of the Acknowledgment of Various Types of Support Systems Over Time (1960-2020). Percentages are column-wise**

These findings differ from a previous study conducted by Yang [16], who found that humanities students tend to express more gratitude towards their relatives, while science students more commonly acknowledge their supervisors. Our results suggest that within the field of LIS, researchers predominantly acknowledge academic support, highlighting the importance of guidance and mentorship from supervisors and committee members in their academic pursuits.

Another set of findings becomes accessible once we look at the various support systems in the context of the gender of both advisor and researcher. We performed a gender analysis of the theses from 1960 onwards[8]. Omitting researcher-advisor pairs where any participants' gender could not be determined, we are left with 603 pairs. The advisors are almost balanced (51.41% female and 48.59% male), whereas the researchers are predominantly female (66.17% female and 33.83% male). Our analysis reveals that advisors, regardless of gender, display a preference for supervising researchers of the same gender (Fig. 2a). This finding raises important questions about gender disparities in research supervision within the field of LIS. It highlights the need for further investigation and intervention to ensure equitable support and opportunities for all researchers.

---

[8] Before 1960, it proved difficult to perform the analysis, because either (i) the gender of either the advisor or the researcher could not be determined, or (ii) the name of the advisor was absent from the metadata, and hence the gender could not be determined.

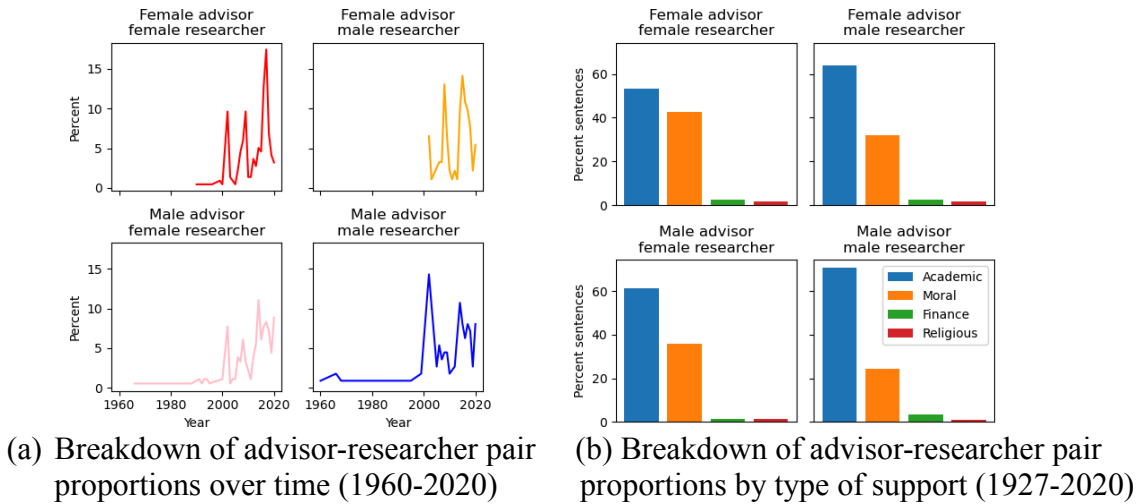

(a) Breakdown of advisor-researcher pair proportions over time (1960-2020)

(b) Breakdown of advisor-researcher pair proportions by type of support (1927-2020)

**Figure 2: Analysis of Acknowledgment Sections by Gender, Broken into the Four Combinations of Female-Female, Female-Male, Male-Female, and Male-Male**

Additionally, we found that female researchers tended to acknowledge both their family/friends and supervisors/committee members more frequently than male researchers in their theses (Fig. 2b). This suggests that female researchers may rely on a broader support network encompassing both personal and academic spheres. Interestingly, the supervisor here seems to make a difference as well: Female researchers acknowledge their academic support system more than their moral support system (60% versus 40%) when they are supervised by a male advisor as opposed to when they are supervised by a female advisor (50% versus 40%). Likewise, male researchers acknowledge their moral support system more when they are supervised by a female supervisor (60% versus 30%) as opposed to when they are supervised by a male supervisor (70% versus 20%). The other two support systems – religious and financial support – do not show strong gender-based differences. Solely male researchers supervised by a male advisor seem to more often acknowledge the financial support they had. Understanding these gender-based differences in acknowledgment patterns provides insights into the complex dynamics of support systems and the unique challenges faced by researchers of different genders.

All in all, the results of this study contribute to our understanding of the support systems utilized by researchers in the field of LIS to complete their theses. These findings shed light on the significance of academic support, the existence of gender disparities in research supervision, and the multifaceted nature of support networks for researchers. These insights can inform efforts to foster inclusive and supportive research environments and promote gender equity in academic pursuits.

### 1.4 Discussion and Conclusion

This study contributes to the existing body of literature on acknowledgment sections by examining ETDs as a previously unexplored source of data. The novelty of this research lies in its focus on investigating the important support systems for male and female researchers in completing their master's or doctoral theses, specifically within the field of Library and Information Science. By utilizing novel computational text analysis methods, we extract features from the acknowledgment sections of ETDs, allowing us to gain insights into the sources of support mentioned.

It is important to note that this study has certain limitations. One limitation is the focus on the field of LIS, which may limit the generalizability of the findings to other academic disciplines. Additionally, the analysis of acknowledgment sections relies on the accuracy of the data extraction process, which may be subject to errors, particularly in cases of OCR inaccuracies. Moreover, there might be errors during the assignment of gender to the advisors' and researchers' names. Finally, even though the metrics for our classifier show strong support for our interpretations, better results might be obtained by annotating more examples, e.g., via utilizing Active Learning [17].

Despite these limitations, this study provides a valuable foundation for further research on acknowledgment sections in ETDs and offers a starting point for exploring support systems in other research domains.

The primary contributions of this study are methodological, and empirical in nature. Firstly, this study enhances our understanding of the context and various uses of expressions found in ETDs' acknowledgment sections. By qualitatively and quantitatively analyzing these sections, we gain insights into the types of support acknowledged. This contributes to a thorough comprehension of the support networks that play a role in researchers' academic achievements. Also, we were able to show gender-based differences in the types of support networks acknowledged, indicating that male and female researchers put different emphasis on various forms of support. Secondly, we could show how novel computational text analysis methods can be fruitfully used to explore large text corpora effectively.

Lastly, this study provides a corpus of annotated sentences for ETDs, serving as a valuable resource for researchers seeking to employ NLP methods themselves. The corpus can be utilized as an example to construct similar datasets for other disciplines, allowing for further exploration and analysis in different research domains. Thus, this study explores ETDs' acknowledgment sections, its examination of support systems in the field of LIS, and the provision of a labeled corpus for future research. By filling the gap in the literature and offering valuable resources, this study makes a significant contribution to the field of acknowledgment analysis and its implications for supporting researchers' academic endeavors.

**Acknowledgments**

The authors would like to thank the organizers of the 2021 Science of Science Summer School where the authors first met and started working on this project.